\documentclass[aps,pra,twocolumn,superscriptaddress]{revtex4}

\usepackage{amsfonts,amssymb,amsmath}
\usepackage{mathtools}
\usepackage[]{graphics,graphicx,epsfig}
\usepackage{amsthm,multirow}
\begin{document}

\title{Quantum information approach to Bose-Einstein condensation of composite bosons}

\author{Su-Yong \surname{Lee}}
\affiliation{Centre for Quantum Technologies, National University of Singapore, 3 Science Drive 2, 117543 Singapore, Singapore}

\author{Jayne \surname{Thompson}}
\affiliation{Centre for Quantum Technologies, National University of Singapore, 3 Science Drive 2, 117543 Singapore, Singapore}

\author{Sadegh \surname{Raeisi}}
\affiliation{Centre for Quantum Technologies, National University of Singapore, 3 Science Drive 2, 117543 Singapore, Singapore}
\affiliation{Institute for Quantum Computing, University of Waterloo, Ontario N2L 3G1, Canada}

\author{Pawe{\l} \surname{Kurzy\'nski}}
\affiliation{Centre for Quantum Technologies, National University of Singapore, 3 Science Drive 2, 117543 Singapore, Singapore}
\affiliation{Faculty of Physics, Adam Mickiewicz University, Umultowska 85, 61-614 Pozna\'{n}, Poland}

\author{Dagomir \surname{Kaszlikowski}}
\email{phykd@nus.edu.sg}
\affiliation{Centre for Quantum Technologies, National University of Singapore, 3 Science Drive 2, 117543 Singapore, Singapore}
\affiliation{Department of Physics, National University of Singapore, 2 Science Drive 3, 117542 Singapore, Singapore}

\date{\today}

\begin{abstract}
We consider composite bosons (cobosons) comprised of two elementary particles, fermions or bosons, in an entangled state. First, we show that 
the effective number of cobosons implies the level of correlation between the two constituent particles. 
For the maximum level of correlation, the effective number of cobosons is the same as the total number of cobosons, which can exhibit the original Bose-Einstein condensation (BEC).
In this context, we study a model of BEC for indistinguishable cobosons with a controllable parameter, i.e., entanglement between the two constituent particles.
We find that bi-fermions behave in a predictable way, i.e., the effective number of the ground state coboson is an increasing function of entanglement between a pair of constituent fermions. Interestingly, bi-bosons exhibit the opposite behaviour -  the effective number of the ground state coboson is a decreasing function of entanglement between a pair of constituent bosons.
\end{abstract}

\maketitle

\section{Introduction}
The idea of Bose-Einstein condensation (BEC) was originally introduced for a uniform, non-interacting gas of elementary bosons \cite{GSS95}. In reality, BEC experiments are conducted using potential traps for gases of bosonic particles, like alkali atoms, atomic hydrogen or metastable helium, that are composite particles made of fermions, and for which inter-particle interactions cannot be neglected \cite{AEMWC95,BSTH95,DMADDKK95}. Alternative BEC scenarios also take into account composite systems, e.g.,  condensation of fermionic pairs in liquid $^3$He \cite{NP90} or excitons (electron-hole pairs) in bulk semiconductors \cite{BBB62,KK65,CN82,EM04,K06}. In addition, these BEC scenarios are closely related to other macroscopic quantum phenomena like superfluidity and superconductivity \cite{NS85,PS02}. 

In many studies the internal structure of composite particles is neglected. On the other hand, it was noted that in some cases this structure plays an important role \cite{NS82,RNPP02,S06,CBD08,CDD09,CSC11,C11}. Therefore, it is interesting to see how BEC can be affected by the internal structure of composite bosonic particles.
Previously BEC was investigated with the interpolation between bosonic and fermionic statistics \cite{AMK03}, and with individual exchanges between the constituent fermions \cite{CS08}. 

In this work we consider a simple model of BEC with composite bosonic particles. In particular, we assume that neither the composite particles nor their constituents interact, such that the internal structure of composite particles is stable and temperature independent. 

Of course, the bound states between constituent particles have to result from their interaction. However, here we assume that once the constituents form a composite particle state, they do not interact anymore. Physically, this may correspond to a dilute gas of composite particles for which energy scales of a binding interaction potential between constituents are much greater than energy scales of the confining trap. As an example, one may think of an atomic hydrogen gas in which ionization temperature is much higher than the standard temperatures required to obtain BEC. Such a simplified model allows us to focus on the fundamental problem of how BEC depends on the internal state of composite particles, while neglecting other physical properties. 

Nowadays, the phenomenology of composite bosons such as excitons, can be explained using the tools developed by quantum information theory \cite{C11}. 
The role of quantum correlations between constituents forming a bound composite particle state can be studied qualitatively and quantitatively using the notion of entanglement. In particular, it was shown that the degree of entanglement between a pair of fermions (bosons) is responsible for their behavior as a single bosonic particle, i.e., only entangled particles behave like a single boson and the more entanglement between them, 
the more their joint bosonic nature is \cite{L05}. 

When even number of particles behaves as boson, the composite particle is called composite boson (coboson). The concept of entanglement was first introduced to coboson with the quantification of the purity of the reduced subsystems \cite{L05}. Then, the analysis of coboson presented that extremizing the purity enhances the bosonic  behavior \cite{COW10,TBM12,RKCSK11,TBM14}. To study on how coboson imitates elementary boson, there were other approaches, such as commutator formalism \cite{C11}, adding and subtracting a single coboson \cite{KRSCK12}, multiple interference of many-coboson \cite{Tichy12}, deformed oscillators \cite{GM12,GM13}, and open quantum system \cite{T13}. Some of the coboson schemes were applied to the investigation of Cooper pairs \cite{PL07}, hydrogen atoms \cite{COW10}, super-bunching effect \cite{TBM13}, coherent states \cite{S13}, and quantum Szilard engine \cite{CK13}. 
Here, we raise the question: how much does coboson BEC deviate from the behavior of a BEC comprised of ideal bosons, using a controllable parameter, i.e., entanglement between the two constituent particles?

Before we start our discussion, let us recall the important results that are relevant to this work. Imagine a pair of distinguishable fermionic or bosonic particles. The system is described by the creation operators $\hat{a}^{\dagger}_k$ and $\hat{b}^{\dagger}_l$, where the indices $k,l = 0,1,\dots, \infty$ label different modes that can be occupied by the two particles. These modes can for example correspond to different energies, or different momentum states. The wave function of the system is of the form
\begin{equation}
\sum_{k,l=0}^{\infty} \alpha_{k,l} \hat{a}^{\dagger}_k \hat{b}^{\dagger}_l |0\rangle,
\end{equation} 
where $\alpha_{k,l}$ is the probability amplitude that particle $a$ is in mode $k$ and particle $b$ is in mode $l$, and $|0\rangle$ is the vacuum state. Using insights from entanglement theory, the mathematical procedure known as the Schmidt decomposition allows us to rewrite the above state as \cite{L05}
\begin{eqnarray}
\sum^{\infty}_{m=0}\sqrt{\lambda_m}\hat{a}^{\dag}_m\hat{b}^{\dag}_m |0\rangle \equiv \hat{c}^{\dag} |0\rangle,
\end{eqnarray}
where the modes labeled by $m$ are superpositions of the previous modes $k$ and $l$ and $\sqrt{\lambda_m}$ are probability amplitudes that both particles occupy mode $m$. Note that despite the fact that $\hat{a}^{\dag}_m$ and $\hat{b}^{\dag}_m$ share the same label, physically these modes might be totally different. What is important is that, the modes labeled by $m$ give rise to the internal structure of a composite particle.

We introduce a composite boson creation operator $\hat{c}^{\dagger}$, that creates a pair of particles. Note that this operator resembles the one for Cooper pairs \cite{C11}. The entanglement between particles is encoded in the amplitudes $\sqrt{\lambda_m}$. In particular, one can introduce a measure of entanglement known as \emph{purity} 
\begin{equation}
P=\sum_{m=0}^{\infty} \lambda_m^2,~~~~0<P \leq 1.
\end{equation}
For $P=1$ the particles are disentangled, whereas in the limit $P \rightarrow 0$ the entanglement between particles goes to infinity. The degree of entanglement can be also expressed via the so called Schmidt number $K=1/P$. Intuitively, $K$ estimates the average number of modes that are taken into account in the internal structure of a composite boson. 

The bosonic properties of $\hat{c}^{\dagger}$ can be studied in many ways. For example, the commutation relation gives $[\hat{c}, \hat{c}^{\dag}]=1+\xi \sum\lambda_m(\hat{a}^{\dag}_m\hat{a}_m+\hat{b}^{\dag}_m\hat{b}_m)$, where $\xi=-1$ if $a$ and $b$ are fermions, or $\xi=+1$ if they are bosons. On the other hand, following the approach in \cite{L05} one may study the ladder properties of this operator
\begin{eqnarray}
&&|n\rangle \equiv \chi_n^{-1/2} \frac{(\hat{c}^{\dag})^n}{\sqrt{n!}} |0\rangle, \nonumber\\
&&\hat{c}|n\rangle=\sqrt{\frac{\chi_{n}}{\chi_{n-1}}}\sqrt{n}|n-1\rangle+|\epsilon_{n}\rangle,
~\langle n-1|\epsilon_{n}\rangle=0,\\
&&\langle \epsilon_{n}|\epsilon_{n}\rangle=1-n\frac{\chi_n}{\chi_{n-1}}+(n-1)\frac{\chi_{n+1}}{\chi_n},\nonumber
\end{eqnarray}
where $|n\rangle$ are states of $n$ composite bosons, parameters $\chi_n$ are normalization factors, such that $\langle n|n \rangle=1$, and $|\epsilon_n\rangle$ are unnormalized states that can result from subtracting a single composite particle from a state $|n\rangle$. The states $|\epsilon_n\rangle$ do not correspond to $n-1$ composite bosons of the same type, but rather to a complicated state of $n-1$ pairs of particles $a$ and $b$. The ladder structure of operators $\hat{c}^{\dagger}$ and $\hat{c}$ starts to approach those of ideal bosons if $\frac{\chi_{n+1}}{\chi_n} \rightarrow 1$ for all $n$. In Ref. \cite{L05,COW10} it has been shown that for a pair of fermions the ratio $\frac{\chi_{n+1}}{\chi_n}$ can be bounded from above and below by the function of entanglement
\begin{equation}
1-nP \leq \frac{\chi_{n+1}}{\chi_n} \leq 1- P. \nonumber
\end{equation}
Then, it has been improved with a tighter upper bound \cite{TBM12}
\begin{equation}
1-nP \leq \frac{\chi_{n+1}}{\chi_n} \leq 1-\frac{nP}{1+(n-1)\sqrt{P}} \leq 1- P. \nonumber
\end{equation}
This result shows that in the limit of large entanglement ($P \ll 1/n$) the pairs of particles behave like elementary bosons. 

To simplify our model, we assume BEC in Gaussian states which are represented by a combination of coherent, thermal, and squeezed states.
Assuming that composite bosons are in a thermal state or in a harmonic trap, we can describe the composite bosons with a Gaussian state. 
Thus, the Gaussian formula of the composite bosons is represented by the following modified operator that is based on the one studied in \cite{L05}
\begin{equation}
\hat{c}^{\dagger}_r = \sum_{m=0}^{\infty} \sqrt{(1-x)x^m} \hat{a}^{\dagger}_{m,r} \hat{b}^{\dagger}_{m,r},
\end{equation}
where the double indices refer to internal ($m$) and to external degrees of freedom ($r$). 
The internal index $m$ may represent their position values. 
In our case $r$ labels the energy levels of the trap in which the BEC takes place. Moreover, as we assumed in the beginning, the internal structure parameters $\lambda_m=(1-x)x^m$ (for $0 \leq x <1$) are independent of $r$. 
The internal structure parameter $\lambda_m$ is equivalent to the coefficient of a two-mode squeezed vacuum (TMSV) state, $|TMSV\rangle=\sum^{\infty}_{m=0}\sqrt{(1-r)r^m}|m\rangle_a|m\rangle_b$, which is a typical two-mode Gaussian state.
The above operator has desirable properties, since it is possible to analytically evaluate the factors $\chi_n$ and one can control the entanglement between constituents $a$ and $b$ via the parameter $x$ \cite{L05}. For $x=0$ the system is separable and in the limit $x\rightarrow 1$ entanglement goes to infinity. In addition
\begin{equation}
0\leq (\frac{\chi_{n+1}}{\chi_n})_F=\frac{x^n(n+1)(1-x)}{(1-x^{n+1})}< 1
\end{equation}
for a pair of fermions \cite{L05} and
\begin{equation}
1< (\frac{\chi_{n+1}}{\chi_n})_B=\frac{(n+1)(1-x)}{(1-x^{n+1})}\leq n+1
\end{equation}
for a pair of bosons \cite{L05}. Finally, the Schmidt number is given by \cite{L05}
\begin{eqnarray}
K=\frac{1+x}{1-x}. 
\end{eqnarray}
The entanglement parameter $x$ is explained with a single hydrogen system in a harmonic trap. The corresponding wave function is given by the product of the ground-state harmonic oscillator and the ground-state electron wave functions, $f(R,b)f(r,a_0)$ \cite{COW10}. $R$ and $r$ represent the positions of the proton and electron, respectively. $b$ is a length parameter characterizing the size of the trap, and $a_0$ is the Bohr radius. Then, the purity of the hydrogen atom is written as $P\sim a_0^3/b^3$ \cite{COW10}. It is related to the entanglement parameter x by the Schmidt number $K=1/P=(1+x)/(1-x)$. Thus, the entanglement parameter $x$ is given in terms of $x\sim[1-a_0^3/b^3]/[1+a_0^3/b^3]$. Therefore, we observe that the entanglement parameter $x$ increases with the length of the size of the trap. It represents that the two distinguishable particles (proton and electron) become more indistinguishable.

The rest of the paper is organized as follows. 
We begin with investigating the meaning of $\langle \hat{c}^{\dag}_r\hat{c}_r\rangle$.
Then, we discuss the BEC of composite bosons made of fermionic pairs. We consider two cases, a potential trap with only two levels and the 3D harmonic potential trap with an infinite number of energy states. Next, we repeat the same for the composite bosons made of bosonic pairs. Finally, we analyze our results in the last section.  
 
\begin{figure}
\centerline{\scalebox{0.35}{\includegraphics[angle=0]{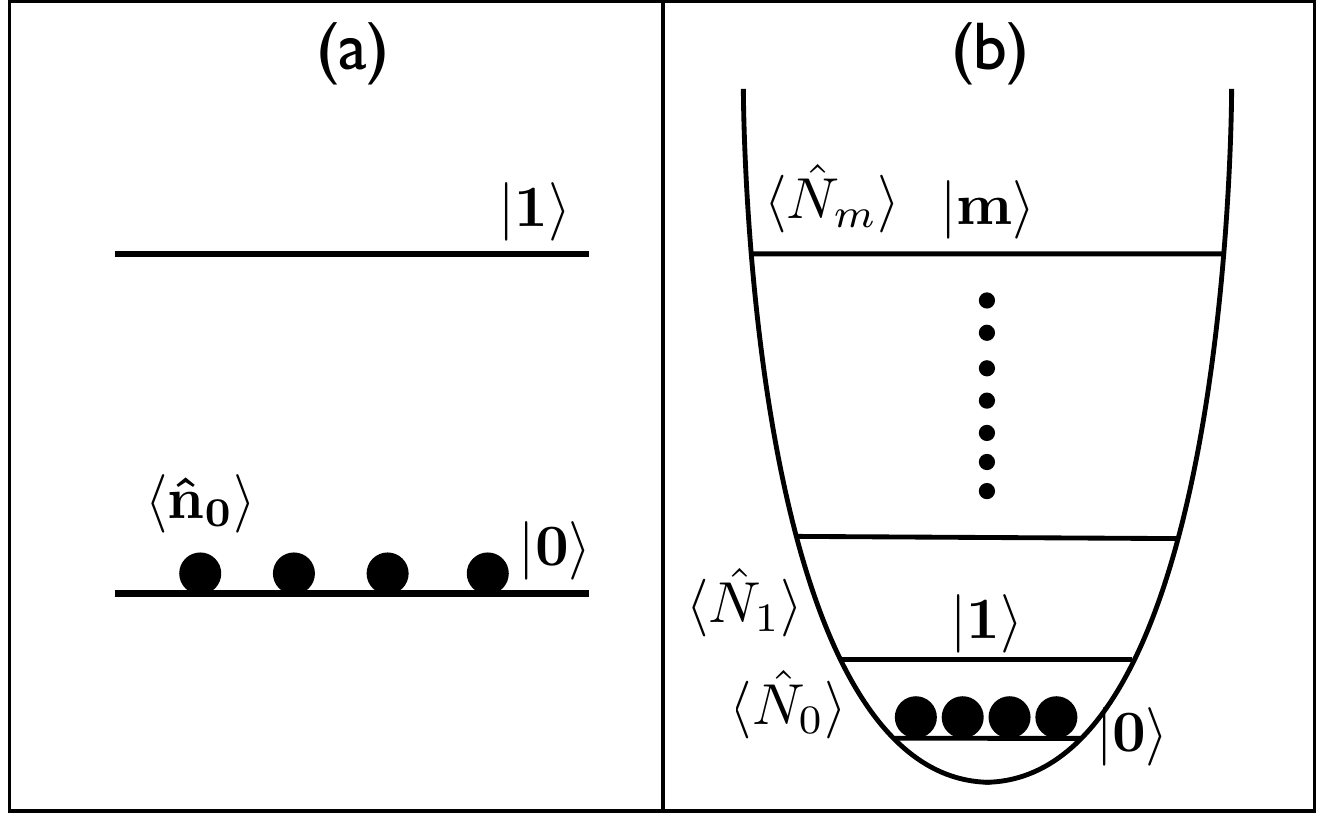}}}
\vspace{-1in}
\caption{BEC using indistinguishable cobosons in (a) a two-level system and (b) a multi-level system.
}
\label{fig:fig1}
\end{figure}

\section{effective number of cobosons}
To count the number of bosons in a specific state, the corresponding bosonic number operator is used. Then, Bose-Einstein condensation (BEC) is investigated by counting the number of bosons in a ground state. Similarly, we consider the coboson effective number operator as a pseudo counting number operator, and observe the phenomena of coboson BEC by counting the pseudo number. In the limit of a high entanglement (low density) between two constituent particles, the coboson effective number operator corresponds to the bosonic number operator, and then the coboson BEC becomes equivalent to the original BEC. 

In the case of two constituent fermions in a multi-level system, the coboson effective number operator is a good indicator for counting the number of cobosons, due to the Pauli exclusion principle. In regimes where the entanglement between two constituent fermions is lower, all the cobosons move to different energy levels so that the coboson effective number operator counts just one coboson. 
In the case of two constituent bosons, the coboson effective number operator cannot be a good indicator for counting the number of cobosons, due to the bunching effect from each constituent boson. In regimes where the entanglement between two constituent bosons is lower, the coboson effective number operator counts more than the total number of cobosons. Thus, we can say that the coboson effective number operator is a good indicator for counting number of bi-fermions in multilevel systems, whereas it exhibits interesting phenomena with bi-bosons.

We look into the meaning of the effective number of cobosons. Using the Eqs. (4) and (5), we evaluate the effective number of cobosons in an $N$ number state $|N\rangle_r$, which represents $N$ cobosons on the r-th energy level, 
\begin{eqnarray}
_r\langle N|\hat{c}^{\dag}_r\hat{c}_r|N\rangle_r=1+(N-1)\frac{\chi_{N+1}}{\chi_{N}}, \nonumber
\end{eqnarray}
where $(\chi_{N+1}/\chi_N)=1$ as the cobosons become ideal bosons.
The ratio $\chi_{N+1}/\chi_N$ is related to the entanglement between the constituent fermions (bosons). 
By many-body effects between two distinguishable fermions \cite{CLT03}, the ratio can be approximated as $\chi_{N+1}/\chi_N\sim 1-Nr^3/w^3$. $r$ is the range of the fermion distribution inside one coboson, $w$ is the width of one coboson, and $N$ is the number of the cobosons. When the width of one coboson increases, the two distinguishable fermions become more indistinguishable. Thus, the entanglement between the constituent fermions increases. Therefore, the ratio $\chi_{N+1}/\chi_N$ increases towards one. In the case of two distinguishable bosons, the ratio can be approximated as $\chi_{N+1}/\chi_N\sim 1+Nr^3/w^3$. With the increasing entanglement between the two distinguishable bosons, the width of one coboson increases so that the ratio $\chi_{N+1}/\chi_N$ gets closer to one.

For pairs of fermions, the normalization ratio $(\chi_{N+1}/\chi_N)_F$ is less than one due to the Pauli exclusion principle between pairs of fermions.
Thus, $_r\langle N|\hat{c}^{\dag}_r\hat{c}_r|N\rangle_r$ is less than the number of cobosons. 
It is explained that the other cobosons move to other energy levels due to the Pauli exclusion principle, which requires the number of cobosons to be less than the number of energy levels.
For pairs of bosons, the normalization ratio $(\chi_{N+1}/\chi_N)_B$ is larger than one due to the bunching effect from the each constituent boson. 
Thus, $_r\langle N|\hat{c}^{\dag}_r\hat{c}_r|N\rangle_r$ is larger than the number of cobosons, which means that the extra cobosons came out of wave nature of the two constituent bosons. It is explained by the second-order correlation functions which represent intensity-intensity correlations \cite{L00}, as below.

Expanding the Eq. (5) in the $_r\langle N|\hat{c}^{\dag}_r\hat{c}_r|N\rangle_r$, the effective number of the cobosons is given by
\begin{eqnarray}
_r\langle N|\hat{c}^{\dag}_r\hat{c}_r|N\rangle_r&=&
\sum^{\infty}_{n=m}\lambda_n\langle \hat{a}^{\dag}_{n,r}\hat{a}_{n,r}\hat{b}^{\dag}_{n,r}\hat{b}_{n,r}\rangle\\
&&+\sum^{\infty}_{n\neq m}\sqrt{\lambda_n\lambda_m}\langle \hat{a}^{\dag}_{n,r}\hat{a}_{m,r}\hat{b}^{\dag}_{n,r}\hat{b}_{m,r}\rangle. \nonumber
\end{eqnarray}
The first term is the sum of unnormalized second-order correlation functions. The second term is, we called, the sum of cross correlation functions.
According to the second-order correlation functions $G^{(2)}$\cite{L00}, Gaussian states of bosons exhibit bunching effect with $1<G^{(2)}<\infty$
whereas fermions exhibit anti-bunching effect with $0\leq G^{(2)}<1$.
Based on the information of the $G^{(2)}$ functions, thus, we can understand the mean of cobosons as follows.
By bunching effect, pairs of bosons can produce $\langle \hat{c}^{\dag}\hat{c}\rangle$ larger than the number of cobosons.
On the other hand, pairs of fermions can reduce the number of cobosons by anti-bunching effect, expelling the other cobosons to other energy levels.

{\it Therefore, the effective number of cobosons represents the level of correlation between the two constituent particles. For the maximum level of correlation, each coboson behaves like a boson. For the weak level of correlation, each constituent particle exhibits its own property so the coboson does not behave like a boson any more. As a controllable parameter for the level of correlation, here, we consider the degree of entanglement between the two constituent particles.}

In the next sections, we deal with the effective number of cobosons in a two-level system and a multi-level one,
using the entanglement between the two constituent particles.


\section{Bi-fermion: a pair of fermions}
We consider indistinguishable cobosons in a two-level system and in a multi-level system, where each coboson is comprised of two fermions (bi-fermion). 
Although the effective number of cobosons in the ground state does not exhibit a BEC phase transition in the two-level system, it is still interesting to compare its thermal behaviour with respect to a two-level system occupied by $N$ cobosons.
We investigate the case in which indistinguishable cobosons are in a Gaussian state, 
such that the normalization ratio of the coboson operator is represented by the parameter $x$ \cite{L05}.
From Eq. (8), $x$ represents the degree of entanglement between a pair of fermions, where
$x=0$ ($x=1$) means that a pair of fermions are separable (maximally entangled).

\subsection{Two-level system: Simplified model}
First we consider a two-level system with a fixed number of $N$ cobosons, see Fig. 1 (a). 
The thermal state of this system reads
\begin{eqnarray}
\rho&=&\frac{1}{Z}\sum^{N}_{n=0}e^{-\beta nE_0}e^{-\beta (N-n)E_1}\nonumber\\
&&|n,N-n\rangle\langle n,N-n|,
\end{eqnarray}
where the total number of cobosons is $N$ and
\begin{eqnarray}
&&|n,N-n\rangle=\frac{(\hat{c}^{\dag}_0)^{n}}{\sqrt{\chi_{n}n!}}\frac{(\hat{c}^{\dag}_1)^{N-n}}{\sqrt{\chi_{N-n}(N-n)!}}|0,0\rangle, \nonumber\\
&&Z=\sum^N_{n=0}e^{-\beta nE_0}e^{-\beta (N-n)E_1},\nonumber
\end{eqnarray}
where $\beta=1/(k_BT)$ and $\chi_{n}$ ($\chi_{N-n}$) is a normalization constant \cite{L05}. $E_0$ and $E_1$ denote the energy levels.
We derive the effective number of cobosons in the ground state as 
\begin{eqnarray}
&&\langle\hat{n}_0\rangle=Tr[\hat{c}^{\dag}_0\hat{c}_0\rho]\\
&&=\frac{1}{Z}\sum^N_{n=0}e^{-\beta nE_0}e^{-\beta (N-n)E_1}[1+(n-1)\frac{\chi_{n+1}}{\chi_{n}}].\nonumber
\end{eqnarray}
Putting $E_0=0$ and $E_1=1$,  the Eq. (11) becomes
\begin{eqnarray}
\langle\hat{n}_0\rangle=\frac{1}{Z}\sum^N_{n=0}e^{-\beta (N-n)}[1+(n-1)\frac{\chi_{n+1}}{\chi_{n}}],
\end{eqnarray}
where the partition function $Z$ is given by $\frac{1-e^{-\beta (N+1)}}{1-e^{-\beta}}$.
For a Gaussian state, the normalization ratio is given by Eq. (6).
When a pair of fermions is not entangled ($x=0$), the effective number of cobosons in the ground state becomes equal to one, regardless of temperature. 
In other words, for $x=0$ there are only two levels that can be occupied by fermionic pairs. Therefore at most two pairs can occupy them due to Pauli exclusion principle.
When a pair of fermions is maximally entangled ($x=1$), the effective number of cobosons in the ground state is given by
\begin{eqnarray}
\langle\hat{n}_0\rangle_{x=1}&=&\frac{1}{1-e^{-\beta(N+1)}}[N-\frac{e^{-\beta}(1-e^{-\beta N})}{1-e^{-\beta}}]\\
&&\xrightarrow{\beta\rightarrow\infty} N.\nonumber
\end{eqnarray}
Hence for maximally entangled fermions the $\langle\hat{n}_0\rangle$ converges to $N$ as temperature tends to zero. In this case the cobosons behave like elementary bosons. The effective number of cobosons is equal to the total mean occupation number of cobosons ($N$) so that all the cobosons occupy the ground state.

For near maximal entanglement ($K\gg N$) between a pair of fermions, we can derive the analytical result by taking the normalization ratio $\chi_{n+1}/\chi_{n}\approx 1-n/K$ \cite{L05}. 
As $T\rightarrow 0$ ($\beta\rightarrow \infty$), the $\langle\hat{n}_0\rangle$ is given by
\begin{widetext}
\begin{eqnarray}
\langle\hat{n}_0\rangle&=&\langle\hat{n}_0\rangle_{x=1}-\frac{N}{K(1-e^{-\beta (1+N)})}[
N-1-\frac{2e^{-\beta}}{1-e^{-\beta}}
+\frac{2e^{-\beta}(1-e^{-\beta N})}{N(1-e^{-\beta})^2}]
\xrightarrow{T\rightarrow 0}N-\frac{N(N-1)}{K}\geq 0,
\end{eqnarray}
\end{widetext}
where the Schmidt number $K$ is represented by the parameter $x$ in Eq. (8). From Eq. (8) and the condition $K\gg N$ , 
the parameter $x$ has the following range $\frac{N-1}{N+1}< x<1$. For $N=100$ we have $0.98< x<1$.
When the Schmidt number $K$ goes to infinity, then the $\langle\hat{n}_0\rangle$ goes to one. 
All the cobosons occupy the ground state energy level $E_0$.

In Fig. 2 we plot the $\langle\hat{n}_0\rangle$ as a function of $T$ against the range of $0.98< x<1$. 
The $\langle\hat{n}_0\rangle$ increases with the degree of entanglement as well as with decreasing temperature. This coincides with the behaviour of an ideal bosonic gas. As $T\rightarrow \infty$, the $\langle\hat{n}_0\rangle$ of cobosons being perfect bosons is saturated with $N/2$. 
When $x$ is slightly less than $1$, the saturation value of cobosons for $T\rightarrow \infty$ can be less than $N/2$.
Note that $\langle\hat{n}_0\rangle<N/2$ is not available in the two-level system because the sum of the effective numbers is less than the total occupation number $N$, 
as $\langle\hat{n}_0\rangle+\langle\hat{n}_1\rangle <N$.
Due to the reason, it is natural to consider BEC in a multi-level system.



\begin{figure}
\vspace{-1in}
\centerline{\scalebox{0.40}{\includegraphics[angle=0]{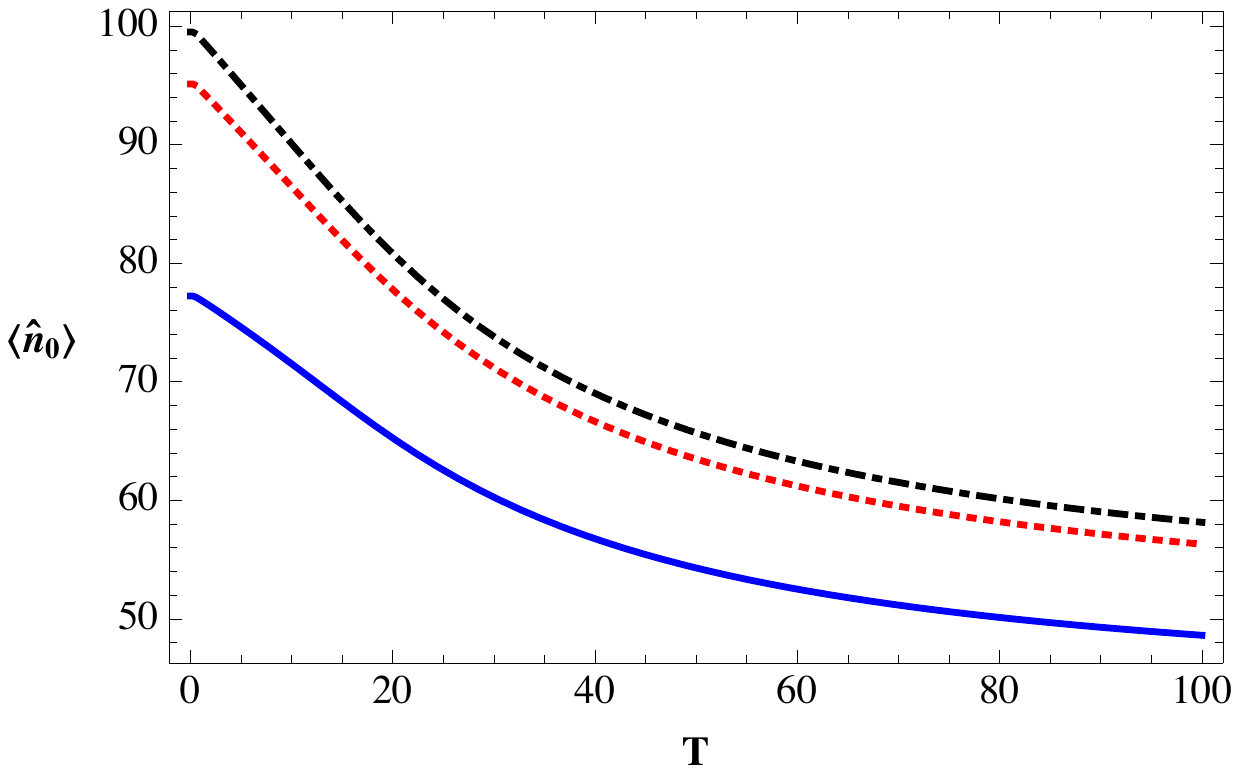}}}
\vspace{-1.2in}
\caption{Effective number of ground state coboson ($N=100$) in a two-level system as a function of $T$: from the bottom to the top ($x=0.995,~0.999,~0.9999$).
The curves are $\langle\hat{n}_0\rangle$ defined in Eq. (12), where the normalization ratio is given by Eq. (6). It indicates that $\langle\hat{n}_0\rangle$ increases with the degree of entanglement ($x$) between a pair of fermions.
}
\label{fig:fig2}
\end{figure}


\subsection{Multi-level system: Realistic model}

\begin{figure}[t]
\centerline{\scalebox{0.3}{\includegraphics[angle=0]{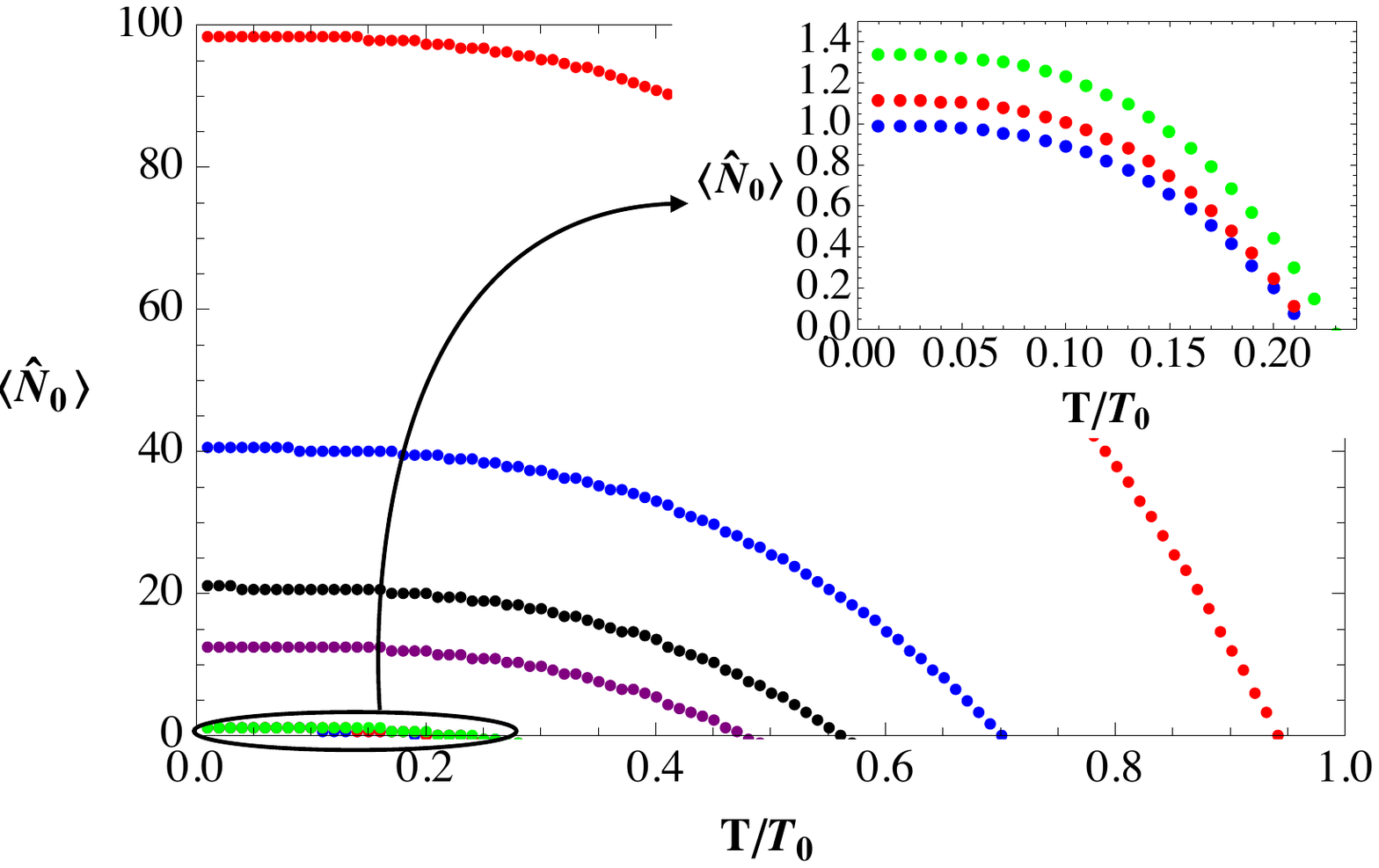}}}
\vspace{-0.4in}
\caption{Effective number of ground state coboson ($\langle\hat{N}\rangle=100$) in a multi-level system as a function of $T/T_0$: from the top to the bottom ($x=0.9999,~0.99,~0.98,~0.97,~0.8,~0.7,~0.001$).  
The small box on the right-side corner represents the $\langle \hat{N}_0\rangle$ for $x=0.8,~0.7,~0.001$.
The curves are $\langle\hat{N}_0\rangle$ defined in Eq. (17), where the normalization ratio is given by Eq. (6). It indicates that $\langle\hat{N}_0\rangle$ increases with the degree of entanglement ($x$) between a pair of fermions. Also, the corresponding transition temperature as the point at which there are no cobosons in the ground state, increases with the degree of entanglement ($x$). 
}
\label{fig:fig3}
\end{figure}

Let us now consider a more realistic physical system consisting of cobosons distributed over the infinitely many energy levels of a 3D isotropic harmonic trap, see Fig. 1 (b). We fix the average number of cobosons to be $\langle \hat{N}\rangle=N$ and describe the system via a grand canonical ensemble with a chemical potential $\mu$. In this paper we do not take the proper thermodynamical limit (such a limit cannot be attained in real experiments) and thus we cannot observe a genuine BEC phase transition. 
In a critical temperature, there is a true transition to BEC which takes the number of particles and the volume to infinity under a constant density. However, it is not possible in real systems. Instead of that, in finite systems, people observed an abrupt transition at some temperature and then a macroscopic number of particles in a lowest energy state. In finite systems, thus, the accumulation point is dealt with a pseudo-critical temperature $T_0$.
Here, we follow Mullin \cite{M97} and investigate the ``pseudo-critical" temperature $T_0$ below which the increase in the chemical potential slows and the number of particles in the ground state begins increasing rapidly. 
We observe that the accumulation point increases with the entanglement parameter $x$, numerically. 

In the grand canonical ensemble, the effective number of the $m$-th energy level $E_m$ and the total effective number are given by
\begin{eqnarray}
\langle \hat{N}_m\rangle&=&\frac{1}{Z_m}\sum^{\infty}_{n=0}e^{-\beta(E_m-\mu)n}[1+(n-1)\frac{\chi_{n+1}}{\chi_n}],\nonumber\\
N&=&\sum_{m=0}^{\infty}\langle \hat{N}_m\rangle,
\end{eqnarray}
where $\hat{N}_m=\hat{c}^{\dag}_m\hat{c}_m$ and $Z_m=(1-e^{-\beta(E_m-\mu)})^{-1}$.
The energy levels in the 3D isotropic harmonic potential are given by $E_m=\hbar\omega(m_x+m_y+m_z+3/2)$, where $m_x,~m_y,~m_z=0,1,2...$
The normalization ratio is given by Eq. (6).
When a pair of fermions is not entangled ($x=0$), the effective number in the ground state is given by
\begin{eqnarray}
\langle \hat{N}_0\rangle=\frac{1}{e^{-\beta\mu}-1}-(1-e^{\beta\mu})\sum^{\infty}_{n=0}e^{\beta\mu n}(n-1)=1,
\end{eqnarray}
where the energy $E_0$ has been taken to be zero.
It exhibits that only one pair of fermions stay on the ground state energy level $E_0$, irrespective of $T$.
Note that we cannot find any temperature dependance of $\langle \hat{N}_0\rangle$ for $x=0$. 
When a pair of fermions is maximally entangled ($x=1$), the effective number of the ground state energy level becomes the same as the Bose-Einstein distribution. 
In this scenario we perfectly recover the conventional Bose-Einstein condensation results.

For all regime of $x$ ($0<x<1$), the effective number of cobosons in the ground state can be numerically estimated using the approximations,
\begin{eqnarray}
&&\langle \hat{N}_0\rangle\approx\langle \hat{N}_0\rangle_{f}-N(\frac{T}{T_0})^3S,\\
&&\langle \hat{N}_0\rangle_{f}= (1-e^{-1/N})\sum^{\infty}_{n=1}e^{-n/N}[1+(n-1)\frac{\chi_{n+1}}{\chi_n}],\nonumber\\
&&S=\sum^{\infty}_{n=1}(\frac{1}{n^3}-\frac{e^{-1/N}}{(1+n)^3})e^{-n/N}[1+(n-1)\frac{\chi_{n+1}}{\chi_n}],\nonumber
\end{eqnarray}
where $S$ approaches $\zeta(3)=\sum^{\infty}_{p=1}\frac{1}{p^3}\approx 1.202$ as $N\rightarrow \infty$ and $x\rightarrow 1$, i.e., 
for an infinite number of maximally entangled cobosons.
The pseudo-critical temperature $T_0$ is given by $\frac{h}{k_B}\sqrt{\frac{U_0}{m}}\rho^{1/3}$ \cite{M97}, where $\rho$ is the average density, $m$ is the mass of a particle, and $U_0$ is the isotropic harmonic potential. 
$\langle \hat{N}_0\rangle_{f}$ represents the effective number in the finite harmonic systems.
The effective number $\langle \hat{N}_0\rangle_{f}$ satisfies the boundary conditions, $\langle\hat{N}_0\rangle_{f}\approx 1$ at $x\sim 0$ (almost no entanglement)
and $\langle\hat{N}_0\rangle_{f}\approx N$ at $x=1$ (maximal entanglement).
For near maximal entanglement between the two constituent fermions, the detailed calculations are given in the Appendix.

We plot the effective number of cobosons in the ground state $\langle \hat{N}_0\rangle$ as a function of $T/T_0$ for different $x$ in Fig. 3.
The $\langle \hat{N}_0\rangle$ increases with decreasing temperature as well as with the degree of entanglement between the two constituent fermions.
At $T/T_0\sim 0$, we find that as the entanglement approaches $0$, the $\langle \hat{N}_0\rangle$ converges to $1$.
It is possible that only one pair of fermions occupy the ground state whereas the rest of pairs of fermions occupy all the different energy levels.
Thus, different from the two-level system, we can observe the $\langle \hat{N}_0\rangle$ in all regime of $x$.
In Fig. 3, we can also see that the transition temperature is an increasing function of entanglement, where we have defined the transition temperature as the point at which there are no cobosons in the ground state. This reflects the fact that the $\langle \hat{N}_0\rangle$ increases with the degree of entanglement.

We can find that our model has some similarities with the references \cite{RNPP02,AMK03}. In the reference \cite{RNPP02}, for a Gaussian state, the maximum occupation number is approximated as $2(W/v)^3$, where $W$ is the width of the one-boson state and $v$ is the width of the fermion distribution inside one boson.
So the maximum occupation number increases with the width of the one-boson state.  In our model, the effective number of bi-fermions in the ground state increases with the entanglement between the two constituent fermions. The entanglement corresponds to the width of the one-boson state, such that the effective number of bi-fermions in the ground state corresponds to the maximum occupation number at $T \rightarrow 0$. 
In the reference \cite{AMK03}, which is about $N$ quons that interpolate between bosonic and fermionic statistics, the condensate depletion is represented by $(N-N_0)/N_0$. $N_0$ is given by $1-q^N/(1-q)$, in which $q=1(-1)$ for boson (fermion).  For $q=1$ (boson), the condensate depletion is equal to $0$ which corresponds to our result that all the cobosons are in the ground state at $T\rightarrow 0$. For $q=-1$ (fermion), the condensate depletion for odd $N$ is equal to $N-1$ which also corresponds to our result that only one pair of fermions stay on the ground state at $T\rightarrow 0$.

As an example, we consider how $T_0$ (pseudo-critical temperature) and $T_c$ (critical temperature) are different in a BEC comprised of atomic hydrogen gas 
for which $T^e_c$ (experimental critical temperature) was observed at $50 \mu K$ \cite{FKWLMKG98}.
Given the density of the hydrogen BEC ($n=1.8\times 10^{20}m^{-3}$), the corresponding theoretical critical temperature in the thermodynamic limit is obtained as $T^t_c=\frac{h^2}{2\pi mk_B}(\frac{n}{\zeta(3/2)})^{2/3}\approx 51\mu K$.
For the finite $N$ systems, the corresponding pseudo-critical temperature is derived by taking the Eq. (11) of \cite{KD96}, $T_c/T_o=1-0.7275/N^{1/3}$. Given $N=10^9$ atoms \cite{FKWLMKG98}, the pseudo-critical temperature increases only $0.073 \%$ from the critical temperature. Thus, the experimental and theoretical pseudo-critical temperatures are $50.0364 \mu K$ and $51.0371 \mu K$, respectively. We find that, for a large number of $N$,  there is almost no difference between the pseudo-critical temperature and the critical temperature.

\section{Bi-boson: a pair of bosons}
We consider cobosons comprised of two bosons (bi-boson).
For a Gaussian state, the normalization ratio is represented by Eq. (7).
Here $x$ parametrizes the degree of entanglement between a pair of bosons.
An example of a coboson is a bi-photon generated by spontaneous parametric down conversion, 
which exhibits composite behavior even if the two photons are spatially separated \cite{L05}.
To keep bi-photons together, we can consider a dye solution which repeatedly absorbs and re-emits photons \cite{KSVW10}.
Previously bi-bosons were considered for super-bunching effect \cite{KRSCK12,TBM13}, and
recently bi-boson systems in an optical lattice were used to observe the correlations in quantum walks \cite{P15}.

\subsection{Two-level system: Simplified model}
We consider a two-level system with a fixed number of $N$ cobosons.
All the formulas used in the previous section are applied here as well - the only difference is the normalization ratio $\chi_{n+1}/\chi_n$.
As we mentioned in Sec. II, due to the bunching effect from the each constituent boson, the effective number of cobosons can be larger than the total mean occupation number of cobosons ($N$) when the degree of entanglement between the two constituent bosons is quantified by a value of $x<1$.
When a pair of bosons is not entangled ($x=0$), 
from Eq. (12) the effective number of cobosons in the ground state is given by
\begin{widetext}
\begin{eqnarray}
\langle \hat{n}_0\rangle_{x=0} &=&\frac{N}{1-e^{-\beta (1+N)}}[N-\frac{2e^{-\beta}}{1-e^{-\beta}}
+\frac{e^{-\beta}(1+e^{-\beta})(1-e^{-\beta N})}{N(1-e^{-\beta})^2}]
\xrightarrow{T\rightarrow 0} N^2,
\end{eqnarray} 
\end{widetext}
where $\beta=1/(k_BT)$. Hence for separable bosons the $\langle \hat{n}_0\rangle_{x=0}$ converges to $N^2$ as temperature tends to zero.
Although the cobosons are no longer behaving like ideal bosons, the dissociated components of each bi-boson pair will 
both independently exhibit bosonic behavior. This causes the $\langle \hat{n}_0\rangle$  to increase as the entanglement between the two constituent bosons decreases. We can see this directly from the formula for $\hat{c}^{\dag}$ in Eq. (5).
At $x=0$ (no entanglement), the coboson operator is represented by the direct product of each component field operator, $\hat{c}^{\dag}=\hat{a}^{\dag}\hat{b}^{\dag}$.
As $T\rightarrow 0$, from Eq. (12), the state of cobosons in the ground state can be described by the coboson number state $|N\rangle$.
So the effective number of the cobosons in the ground state is given by
\begin{eqnarray}
\langle N|\hat{c}^{\dag}\hat{c}|N\rangle=\langle N_a,N_b|\hat{a}^{\dag}\hat{a}\hat{b}^{\dag}\hat{b}|N_a,N_b\rangle=N^2,
\end{eqnarray}
where $a$ and $b$ represent different modes. 
Note that the effective number of bi-bosons is not the same as the mean occupation number of the dissociated components of bi-bosons ($2N$) at $x=0$.
It can be explained that the enormous value $N^2$ comes out of the sum of the correlation functions in Eq. (9), where the correlations functions can exhibit super-bunching effects by wave nature.
When a pair of bosons is maximally entangled ($x=1$), the $\langle \hat{n}_0\rangle$ converges to $N$ as temperature goes to zero.

For near maximal entanglement ($K\gg N$) between a pair of bosons, we can make the approximation, $\chi_{n+1}/\chi_n\approx 1+n/K$ \cite{L05}.
As $T\rightarrow 0 $, the $\langle \hat{n}_0\rangle$ approaches
\begin{widetext}
\begin{eqnarray}
\langle\hat{n}_0\rangle=\langle\hat{n}_0\rangle_{x=1}+\frac{N}{K(1-e^{-\beta (1+N)})}[
N+1-\frac{2e^{-\beta}}{1-e^{-\beta}}
+\frac{2e^{-2\beta}(1-e^{-\beta N})}{N(1-e^{-\beta})^2}]
\xrightarrow{T\rightarrow 0}N+\frac{N(N+1)}{K}\geq N,
\end{eqnarray}
\end{widetext}
where $\langle\hat{n}_0\rangle_{x=1}$ is given by Eq. (13).
If the Schmidt number $K$ goes to infinity, then the $\langle \hat{n}_0\rangle$ goes to $N$.
For all regimes of $x$ ($0<x<1$), we plot the $\langle \hat{n}_0\rangle$ as a function $T$ in Fig. 4.
The $\langle \hat{n}_0\rangle$ decreases with increasing temperature as well as with the degree of entanglement between the two constituent bosons.
At $T\sim0$, the $\langle \hat{n}_0\rangle$ is maximized as a decreasing function of entanglement which ranges from $N^2$ to $N$.
In contrast to cobosons comprised of fermions, therefore, the $\langle \hat{n}_0\rangle$ decreases with entanglement between the two constituent bosons,
due to the super-bunching effects between the two constituent bosons.

\begin{figure}
\centerline{\scalebox{0.3}{\includegraphics[angle=0]{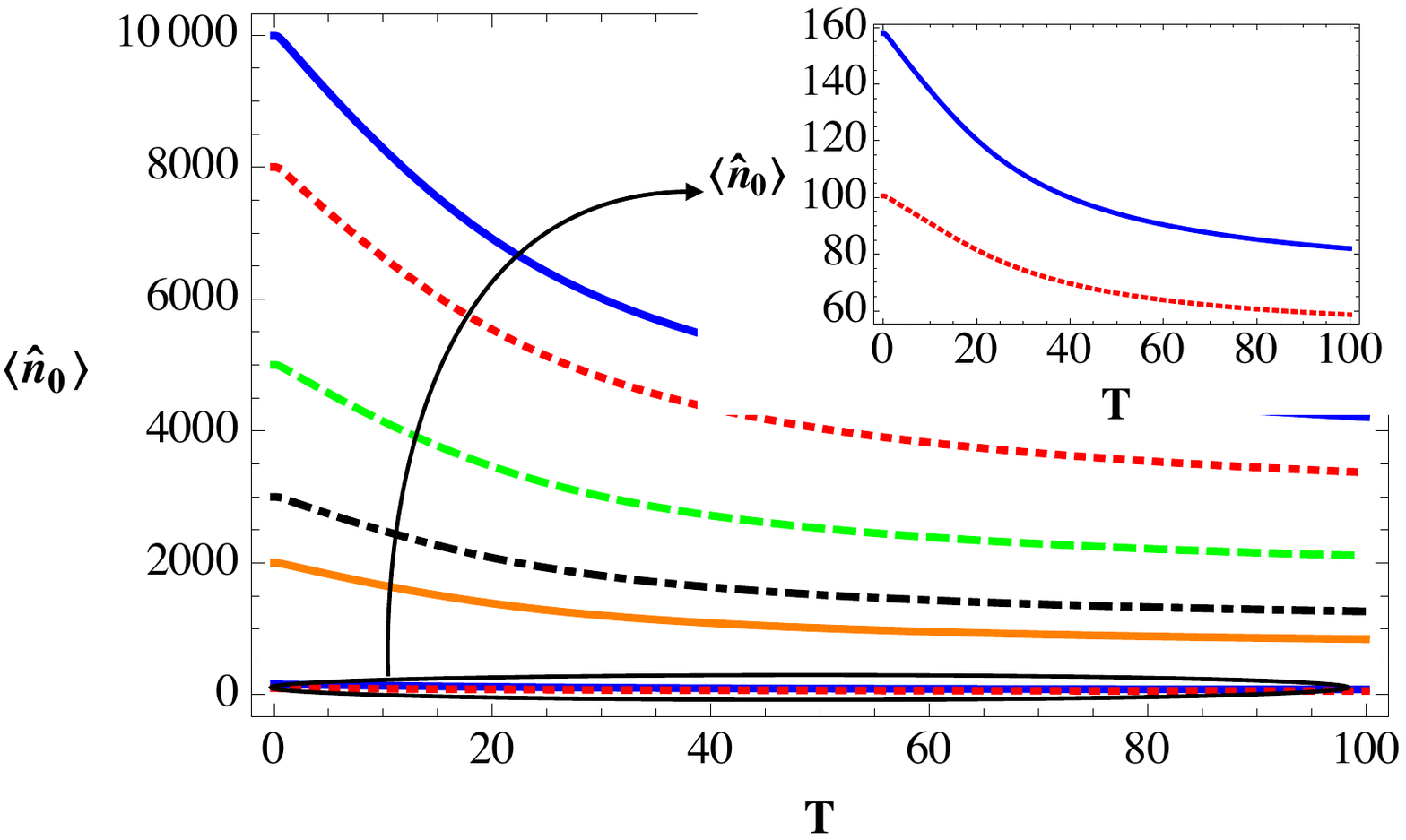}}}
\vspace{-0.5in}
\caption{Effective number of ground state coboson ($N=100$) in a two-level system as a function of $T/T_0$: from the bottom to the top ($x=0.9999,~0.99,~0.8,~0.7,~0.5,~0.2,~0.001$).  
The small box on the right-side corner represents the condensate fraction for $x=0.9999,~0.99$.
Here $x$ represents the degree of entanglement for a pair of bosons.
The curves are $\langle\hat{n}_0\rangle$ defined in Eq. (12), where the normalization ratio is given by Eq. (7). It indicates that $\langle\hat{n}_0\rangle$ decreases with the degree of entanglement ($x$) between a pair of bosons.
}
\label{fig:fig4}
\end{figure}

\subsection{Multi-level system: Realistic model}
We consider a 3D isotropic harmonic trap which contains an average of $N$ cobosons.
When a pair of bosons is not entangled ($x=0$), the effective number of the ground state in Eq. (15) is given by
\begin{eqnarray}
\langle \hat{N}_0\rangle=\frac{1}{Z_0}\sum^{\infty}_{n=0}e^{-\beta(E_0-\mu)n}n^2=\frac{z(1+z)}{(1-z)^2},
\end{eqnarray} 
where $z=\exp(\beta\mu)$ is the fugacity and $E_0$ has been taken to be zero. Compared with the Bose-Einstein (BE) distribution where $N_0=z/(1-z)$, 
 the $\langle \hat{N}_0\rangle$  is always greater than the BE distribution one. 
When a pair of bosons is maximally entangled ($x=1$),  the $\langle \hat{N}_0\rangle$ is the same as for BE distribution.
This reaffirms the potential for BEC of bi-bosons.

In Fig. 5 we plot the $\langle \hat{N}_0\rangle$ as a function of $T/T_0$, for bi-bosons exhibiting a range of entanglement values.
The $\langle \hat{N}_0\rangle$ decreases with the degree of entanglement between the two constituent bosons.
As bi-bosons become less entangled they behave more like a system of two independent bosons.
Hence at $T/T_0\sim 0$, the $\langle \hat{N}_0\rangle$ is maximized as a decreasing function of entanglement.
Using Eq. (20), we derive the maximum condensate fraction at $x\sim 0$ as
\begin{eqnarray}
\langle \hat{N}_0\rangle_{x=0}=(1-e^{-1/N})\sum^{\infty}_{n=1}e^{-n/N}n^2 \approx 2N^2,
\end{eqnarray}
where $N$ is sufficiently large.
In Fig. 5, we can also see that the transition temperature decreases with increasing entanglement.
This reflects the fact that the $\langle \hat{N}_0\rangle$ decreases with the degree of entanglement.
Therefore, similarly to the two-level system, we observe that the $\langle \hat{N}_0\rangle$ decreases as a function of entanglement between the two constituent bosons.

We have never experimentally observed the phenomenon of bi-boson BEC, but a BEC experiment has been conducted using photons in an optical micro-cavity \cite{KSVW10}. Based on the techniques used to create a BEC from photons, we look forward to observing future bi-photon condensates in optical cavities.

\begin{figure}
\centerline{\scalebox{0.3}{\includegraphics[angle=0]{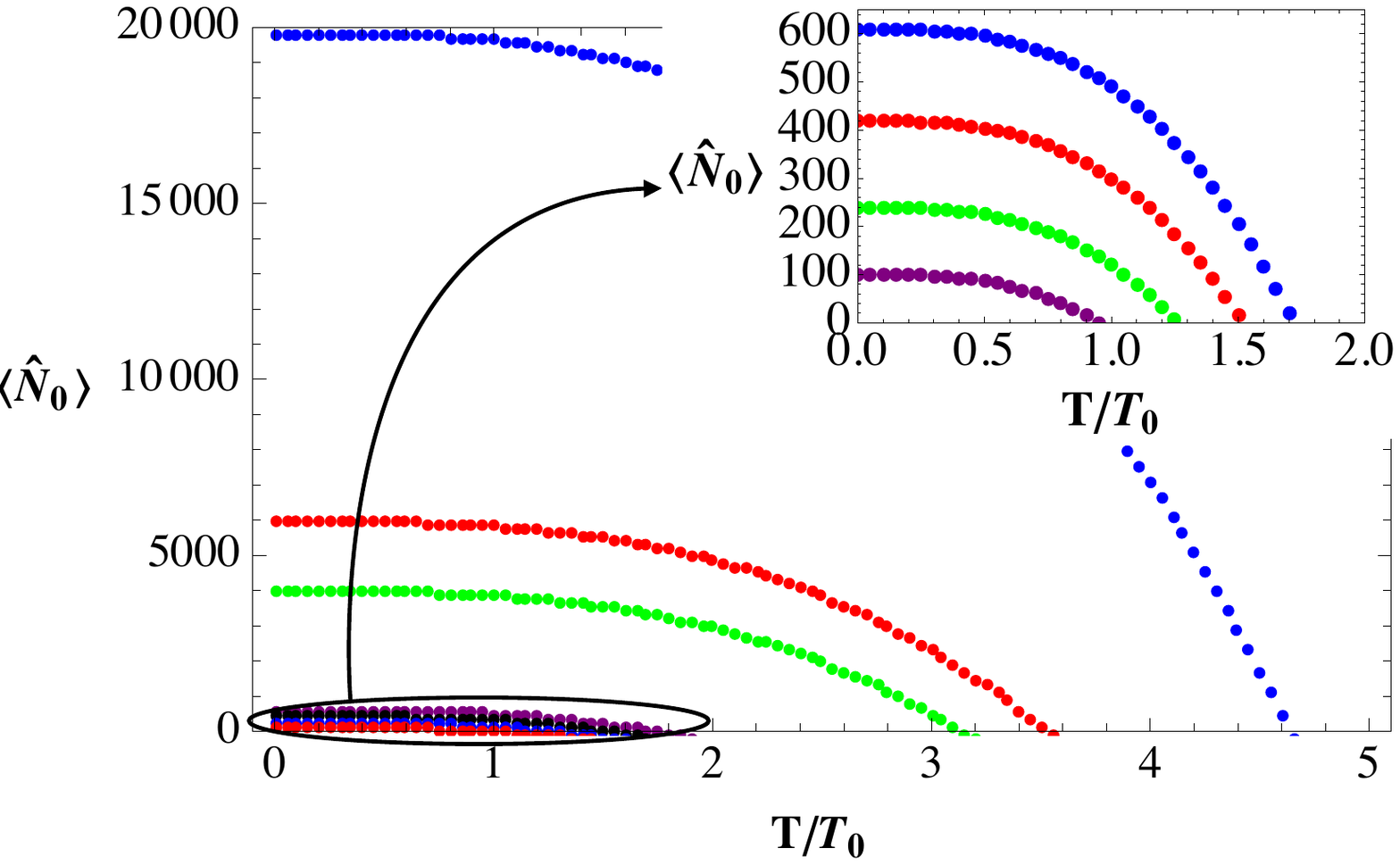}}}
\vspace{-0.42in}
\caption{Effective number of ground state coboson ($\langle\hat{N}\rangle=100$) in a multi-level system as a function of $T/T_0$: from the bottom to the top ($x=0.9999,~0.99,~0.98,~0.97,~0.8,~0.7,~0.001$).  
The small box on the right-side corner represents the condensate fraction for $x=0.9999,~0.99,~0.98,~0.97$.
Here $x$ represents the degree of entanglement for a pair of bosons. 
The curves are $\langle\hat{N}_0\rangle$ defined in Eq. (17), where the normalization ratio is given by Eq. (7). It indicates that $\langle\hat{N}_0\rangle$ decreases with the degree of entanglement ($x$) between a pair of bosons. Also, the corresponding transition temperature as the point at which there are no cobosons in the ground state, decreases with the degree of entanglement ($x$).
}
\label{fig:fig5}
\end{figure}

\section{Conclusion}
It is known that the effective number of cobosons is related to the level of correlation between the two constituent particles. 
For the maximum level of correlation, the effective number of cobosons is the same as the total number of cobosons.
For the weak level of correlation, the effective number of cobosons is smaller (larger) than the total number of cobosons while each constituent fermion (boson) exhibits its own property.

Then we studied how much the coboson BEC deviates from the behavior of a BEC comprised of ideal bosons, using a controllable parameter, i.e., entanglement between the two constituent particles. We specifically considered bi-fermions trapped in a 3D isotropic harmonic system.
By the Pauli exclusion principle between bi-fermions, we found that the effective number of bi-fermions can be smaller than the total number of bi-fermions, regardless of system. Thus we demonstrated that the effective number of bi-fermions in the ground state 
 increases with the degree of entanglement between a pair of fermions.
{\it Correspondingly, we found that the transition temperature for the 3D isotropic harmonic system, i.e., the temperature at which all the bi-fermions moved to the excited states, increased with increasing entanglement.}

Furthermore, we discussed coboson BEC, where each coboson is a bi-boson.
Due to the bunching effect between the constituent bosons, the effective number of bi-bosons can be greater than the total number of bi-bosons.
Thus it was shown that the effective number of bi-bosons in the ground state decreases with the degree of entanglement between a pair of bosons.
{\it Correspondingly, the transition temperature for the 3D isotropic harmonic system decreased with increasing entanglement.
When the entanglement between a pair of bosons becomes sufficiently small, the bi-boson pairs are dissociated, increasing the bunching effect in the effective number of bi-bosons. Consequently the coboson operator is represented by the direct product of each component field operator.}

As further work, it would be interesting to study the actual occupation number for cobosons and how entanglement between a pair of fermions (bosons) could affect super-radiance in coboson BECs.

\begin{acknowledgments}
S.Y.L. thanks C. Noh and T.K.C. Bobby for useful suggestions.
This work was supported by the National Research Foundation and Ministry of Education in Singapore and the Academic Research Fund Tier 3 MOE2012-T3-1-009.
S.R. was supported by Canada's NSERC, MPrime, CIFAR, and CFI and IQC.
\end{acknowledgments}

\appendix
\section*{Appendix: Near maximal entanglement between the two constituent fermions in the 3D isotropic harmonic potential}
For near maximal entanglement between a pair of fermions [$x=1-\delta$ ($\delta \ll 1$)], 
we can use the approximation $\chi_{n+1}/\chi_n\sim x^n$
to analytically derive  the effective number for each energy level:
\begin{eqnarray}
\langle \hat{N}_m\rangle=\frac{(1+2\delta)}{e^{\beta(E_m-\mu)}-1}-\frac{e^{\beta(E_m-\mu)}2\delta}{(e^{\beta(E_m-\mu)}-1)^2}. \nonumber
\end{eqnarray} 
The total effective number of cobosons reads
\begin{eqnarray}
N =\sum^{\infty}_{m}\frac{1}{e^{\beta(E_m-\mu)}-1}[1+2\delta(1-\frac{ e^{\beta(E_m-\mu)}}{e^{\beta(E_m-\mu)}-1})], \nonumber
\end{eqnarray}
where $m=m_x,m_y,m_z$.

Using the relation $\beta\hbar\omega=T_0/(TN^{1/3})$ \cite{M97}, the total effective number of cobosons is given by
\begin{widetext}
\begin{eqnarray}
N=\sum^{\infty}_{p=0}[(1+2\delta)\frac{\frac{1}{2}p^2+\frac{3}{2}p+1}{e^{T_0/(TN^{1/3})p+\alpha}-1}
-2\delta\frac{(\frac{1}{2}p^2+\frac{3}{2}p+1)e^{T_0/(TN^{1/3})p+\alpha}}{(e^{T_0/(TN^{1/3})p+\alpha}-1)^2}],
\end{eqnarray}
\end{widetext}
where $\alpha= 3T_0/(2TN^{1/3})-\beta\mu$ and $p=m_x+m_y+m_z$. 
The coefficient, $\frac{1}{2}p^2+\frac{3}{2}p+1$, originated from the energy level degeneracy, i.e., $p=m_x+m_y+m_z$.
The first term of Eq. (23) is given by \cite{M97}
\begin{eqnarray}
\langle \hat{N}\rangle_1&\equiv& \langle\hat{N}_0\rangle_1+(\frac{T}{T_0})^3NF_3(\gamma)+\frac{5}{2}(\frac{T}{T_0})^2N^{2/3}F_2(\gamma)\nonumber\\
&&+3(\frac{T}{T_0})N^{1/3}F_1(\gamma),\nonumber
\end{eqnarray}
where $\langle\hat{N}_0\rangle_1=1/(e^{\alpha}-1)$ and $\gamma=\alpha+T_0/(TN^{1/3})$. 
$F_{l}(\gamma)\equiv \frac{1}{(l-1)!}\int^{\infty}_{0}du\frac{u^{l-1}}{e^{u+\gamma}-1}=\sum^{\infty}_{p=1}\frac{e^{\gamma p}}{p^l}$ are the Bose integrals.
The second term of Eq. (23) is derived as
\begin{eqnarray}
\langle \hat{N}\rangle_2&\equiv&\frac{e^{\alpha}}{(e^{\alpha}-1)^2}+\sum^{\infty}_{p=1}\frac{(\frac{1}{2}p^2+\frac{3}{2}p+1)e^{T_0/(TN^{1/3})p+\alpha}}{(e^{T_0/(TN^{1/3})p+\alpha}-1)^2}\nonumber\\
&=&\langle\hat{N}_0\rangle_2+\sum^{\infty}_{q=0}\frac{(\frac{1}{2}q^2+\frac{5}{2}q+3)e^{T_0/(TN^{1/3})q+\gamma}}{(e^{T_0/(TN^{1/3})q+\gamma}-1)^2},\nonumber
\end{eqnarray}
where $q=p-1$ to sum up from $0$, and $\langle\hat{N}_0\rangle_2=e^{\alpha}/(e^{\alpha}-1)^2$.
For small temperature, $\alpha$ is quantified as $1/N$ in the finite $N$ harmonic systems \cite{M97}.
Since the states get closely spaced in large $N$ \cite{M97}, the summation can be replaced by an integral such that
\begin{eqnarray}
\langle \hat{N}\rangle_2&=&\langle\hat{N}_0\rangle_2+\frac{1}{2}(\frac{T}{T_0})^3N\int^{\infty}_{0}du\frac{u^2e^{u+\gamma}}{(e^{u+\gamma}-1)^2}\nonumber\\
&&+\frac{5}{2}(\frac{T}{T_0})^2N^{2/3}\int^{\infty}_{0}du\frac{ue^{u+\gamma}}{(e^{u+\gamma}-1)^2}\nonumber\\
&&+3(\frac{T}{T_0})N^{1/3}\int^{\infty}_{0}du\frac{e^{u+\gamma}}{(e^{u+\gamma}-1)^2}.\nonumber
\end{eqnarray}
Then, it is represented by 
\begin{eqnarray}
\langle \hat{N}\rangle_2&=&\langle\hat{N}_0\rangle_2-(\frac{T}{T_0})^3NF_2(\gamma)-\frac{5}{2}(\frac{T}{T_0})^2N^{2/3}F_1(\gamma)\nonumber\\
&&-3(\frac{T}{T_0})N^{1/3}\frac{dF_1(\gamma)}{d\gamma}.\nonumber
\end{eqnarray}
In the thermodynamic limit, which requires to increase the volume of the system and the number of particles while the average density is fixed, the Eq. (23) is derived as
\begin{eqnarray}
N&=&\langle \hat{N}\rangle_1+\langle \hat{N}\rangle_2\nonumber\\
&\approx & \langle \hat{N}_0\rangle+N(\frac{T}{T_0})^3[\zeta(3)+2\delta(\zeta(3)+\zeta(2))],\nonumber
\end{eqnarray}
where we considered $F_{l}(\gamma)\approx \zeta(l)$ for small $\gamma$ \cite{M97}, 
and $\langle \hat{N}_0\rangle=(1+2\delta)\langle \hat{N}_0\rangle_1-2\delta\langle \hat{N}_0\rangle_2$. 
$\zeta(l)=\sum^{\infty}_{p=1}1/p^l$ is the Riemann zeta function.


\begin{thebibliography}{99}

\bibitem{GSS95} A. Griffin, D.W. Snoke, and S. Stringari, {\it Bose-Einstein Condensation} (Cambridge Univ. Press, Cambridge, 1995).

\bibitem{AEMWC95} M.H. Anderson, J.R. Ensher, M.R. Matthews, C.E. Wieman, and E.A. Cornell, Science \textbf{269}, 198 (1995).

\bibitem{BSTH95} C.C. Bradley, C.A. Sackett, J.J. Tollett, and R.G. Hulet, \prl \textbf{75}, 1687 (1995).

\bibitem{DMADDKK95} K.B. Davis, M.-O. Mewes, M.R. Andrews, N.J. van Druten, D.S. Durfee, D.M. Kurn, and W. Ketterle, \prl \textbf{75}, 3969 (1995).

\bibitem{NP90} P. Nozieres and D. Pines, {\it The Theory of Quantum Liquids, Vol. II} (Addison-Wesley,Eading, Mass., 1990).

\bibitem{BBB62} J.M. Blatt, K.W. B\"oer, and W. Brandt, Phys. Rev. \textbf{126}, 1691 (1962).

\bibitem{KK65} L.V. Keldysh and Y.V. Kopaev, Sov. Phys. Solid. State \textbf{6}, 2219 (1965).

\bibitem{CN82} C. Comte and P. Nozieres, J. Phisique \textbf{43}, 1069 (1982); ibid. \textbf{43}, 1083 (1982).

\bibitem{EM04} J.P. Eisenstein and A.H. MacDonald, Nature \textbf{432}, 691 (2004).

\bibitem{K06} J. Kasprzak et al., Nature \textbf{443}, 409 (2006).

\bibitem{NS85} P. Nozieres and S. Schmitt-Rink, J. Low. Temp. Phys. \textbf{59}, 195 (1985).

\bibitem{PS02} C.J. Pethick and H. Smith, {\it Bose-Einstein Condensation in Dilute Gases} (Cambridge University Press, 2002).

\bibitem{NS82} P. Nozieres and D. Saint James, J. Phys. France \textbf{43}, 1133 (1982).

\bibitem{RNPP02} S. Rombouts, D.V. Neck, K. Peirs, and L. Pollet, Mod. Phys. Lett. A \textbf{17}, 1899 (2002).

\bibitem{S06} P. Sancho, J. Phys. A: Math. Theor. \textbf{39}, 12525 (2006).

\bibitem{CBD08} M. Combescot, O. Betbeder-Matibet, and F. Dubin, Phys. Rep. \textbf{463}, 215 (2008).

\bibitem{CDD09} M. Combescot, F. Dubin, and M. A. Dupertuis, \pra \textbf{80}, 013612 (2009).

\bibitem{CSC11} M. Combescot, S.-Y. Shiau, and Y.-C. Chang, \prl \textbf{106}, 206403 (2011).

\bibitem{C11} M. Combescot, Europhys. Lett. \textbf{96}, 60002 (2011).

\bibitem{AMK03} S.S. Avancini, J.R. Marinelli, and G. Krein,  J. Phys. A: Math. Theor. \textbf{36}, 260403 (2003).

\bibitem{CS08} M. Combescot and D.W. Snoke, \prb \textbf{78}, 144303 (2008).

\bibitem{L05} C.K. Law, \pra \textbf{71}, 034306 (2005).

\bibitem{COW10} C. Chudzicki, O. Oke, and W.K. Wootters, \prl \textbf{104}, 070402 (2010).

\bibitem{TBM12} M.C. Tichy, P.A. Bouvrie, and K. M\o lmer, \pra \textbf{86}, 042317 (2012).

\bibitem{RKCSK11} R. Ramanathan, P. Kurzy\'nski, T.K. Chuan, M.F. Santos, and D. Kaszlikowski, \pra \textbf{84}, 034304 (2011).

\bibitem{TBM14} M.C. Tichy, P.A. Bouvrie, and K. M\o lmer,  Appl. Phys. B \textbf{117}, 785 (2014).


\bibitem{KRSCK12} P. Kurzy\'nski, R. Ramanathan, A. Soeda, T.K. Chuan, and D. Kaszlikowski, New J. Phys. \textbf{14}, 093047 (2012).

\bibitem{Tichy12} M.C. Tichy, P.A. Bouvrie, and K. M\o lmer, \prl \textbf{109}, 260403 (2012).

\bibitem{GM12} A.M. Gavrilik, and Y.A. Mishchenko, Phys. Lett. A \textbf{376}, 1596 (2012).

\bibitem{GM13} A.M. Gavrilik, and Y.A. Mishchenko, J. Phys. A: Math. Theor. \textbf{46}, 145301 (2013).

\bibitem{T13} A. Thilagam, J. Math. Chem. \textbf{51}, 1897 (2013).

\bibitem{PL07} Y. Pong and C.K. Law, \pra \textbf{75}, 043613 (2007).

\bibitem{TBM13} M.C. Tichy, P.A. Bouvrie, and K. M\o lmer, \pra \textbf{88}, 061602 (R) (2013).

\bibitem{S13} S.-Y. Lee, J. Thompson, P. Kurzy\'nski,  A. Soeda, and D. Kaszlikowski, \pra \textbf{88}, 063602 (2013).

\bibitem{CK13} T.K. Chuan, and D. Kaszlikowski, arXiv:1308.1525. 


\bibitem{CLT03} M. Combescot, X. Leyronas, and C. Tanguy, Eur. Phys. J. B 31, 17-24 (2003).

\bibitem{L00} R. Loudon, {\it The Quantum Theory of Light} (Oxford Univ. Press, Oxford, 2000).

\bibitem{KD96} W. Ketterle and N.J. van Druten, \pra \textbf{54}, 656 (1996).

\bibitem{M97} W.J. Mullin, J. Low Temp. Phys. \textbf{106}, 615 (1997).

\bibitem{FKWLMKG98} D.G. Fried, T.C. Killian, L. Willmann, D. Landhuis, S.C. Moss, D. Kleppner, and T.J. Greytak, \prl \textbf{81}, 3811 (1998).

\bibitem{KSVW10} J. Klaers, J. Schmitt, F. Vewinger, and M. Weitz, Nature \textbf{468}, 545 (2010).

\bibitem{P15} P.M. Preiss et al., Science \textbf{347}, 1229 (2015).
















 



















\end{thebibliography}
\end{document}